\documentclass[aps,prc,numerical,superscriptaddress,floatfix,reprint]{revtex4-1}
\usepackage{amsmath,amssymb,mathrsfs,bm} 
\usepackage{url}
\usepackage{hyperref}
\hypersetup{
colorlinks=true,
linkcolor=blue,
anchorcolor =red,
citecolor=blue,
filecolor = red,
urlcolor=blue,
pdfauthor=author}
\usepackage{txfonts}
\usepackage[utf8]{inputenc}
\usepackage{amsmath}
\usepackage{graphicx}
\def\av<#1>{\left\langle\,#1\,\right\rangle}
\def\ev<#1>{\left\langle\,#1\,\right\rangle_{\rm{ev}}}

\begin{document}

\title{Heavy-Quark Dynamics from $D^{0}$-Meson Anisotropy-Response Scaling Violations}

\author{ Roy~A.~Lacey}
\email[E-mail: ]{Roy.Lacey@Stonybrook.edu}
\affiliation{Department of Chemistry, 
Stony Brook University, \\
Stony Brook, NY, 11794-3400, USA}
%
%

\date{\today}
\begin{abstract}
A species-resolved anisotropy-response scaling framework is used
to investigate heavy-quark dynamics through the elliptic ($v_2$)
and triangular ($v_3$) azimuthal anisotropies of $D^0$ mesons.
The independently constrained light-flavor response provides a
direct null test: the transformed $D^0$ harmonics exhibit approximate
inter-harmonic closure but remain systematically displaced from
the kaon-anchored scaling function, revealing a charm scaling
violation. The common response is recovered through a charm-specific
attenuation rescaling, $k_c$, together with a small low-$p_T$
hadronic-rescattering correction, $\zeta_{\rm hs}\simeq0.06$, while
retaining the established light-flavor prescription.
In Pb+Pb collisions at $\sqrt{s_{NN}}=5.02$~TeV,
$k_c$ decreases from approximately $2.0$ in central and mid-central
collisions to $1.27$ in the 30--50\% interval. For $D^0$ $v_2$,
ultra-central p+Pb collisions at $\sqrt{s_{NN}}=8.16$~TeV
require $k_c\simeq1.5$, whereas Au+Au collisions at
$\sqrt{s_{NN}}=200$~GeV are consistent with $k_c\simeq1$.
The transition momentum exhibits a distinct system dependence:
$D^0$ and light mesons show comparable
$p_{T,\rm tr}\simeq4.5$--$5.0$~GeV/$c$ in Pb+Pb, consistent
with the common large-system prescription also used in Au+Au,
whereas p+Pb $D^0$ measurements are consistent with the lower
light-meson scale of approximately $1.6$~GeV/$c$.
These complementary constraints on attenuation strength and
transition momentum provide a differential benchmark for
heavy-quark transport and medium evolution.
\end{abstract}

%

\maketitle

%
%
Heavy quarks provide distinctive probes of the strongly interacting
matter produced in relativistic nuclear collisions because they are
produced predominantly in initial hard-scattering processes and
subsequently propagate through the evolving medium. Their momentum
distributions are modified by diffusion and drag, collisional and
radiative energy loss, collective expansion, and hadronization
\cite{Prino:2016cni,Rapp:2018qla}. A central objective of heavy-flavor
measurements is therefore to determine how much of the observed
response reflects the evolving bulk medium and how much carries
additional information specific to heavy-quark transport
\cite{Cao:2018ews}.

A key observable is the azimuthal anisotropy of produced particles,
conventionally expressed as
\begin{equation}
\frac{dN}{d\phi}\propto
1+2\sum_{n=1}^{\infty}v_n
\cos[n(\phi-\Psi_n)],
\label{eq:vn_definition}
\end{equation}
where $v_n$ denotes the $n$th-order anisotropy coefficient and
$\Psi_n$ the corresponding symmetry-plane angle. In particular,
$v_2$ and $v_3$ characterize elliptic and triangular anisotropies,
respectively. Substantial $D^0$ anisotropies have been measured at
both RHIC and the LHC 
\cite{STAR:2017kkh,ALICE:2017pbx,CMS:2017vhp,CMS:2020bnz}, including 
simultaneous measurements of prompt $D^0$ $v_2$ and $v_3$ in Pb+Pb collisions
\cite{CMS:2020bnz}. These measurements establish a significant charm 
anisotropy but do not, by themselves, distinguish the response shared 
with the light-flavor sector from that associated specifically with charm
transport.

This separation is essential because many of the mechanisms
that generate $D^0$ anisotropy also influence light-flavor
hadrons. Initial-state geometry, collective expansion,
viscous attenuation, path-length-dependent energy loss,
and hadronization contribute to both sectors \cite{He:2022ywp}. 
The central question is therefore whether charm follows the same
anisotropy response as light flavor, rather than simply
whether it develops substantial $v_n$.

Species-resolved anisotropy-response scaling provides an
experimentally constrained basis for this comparison
\cite{Lacey:2024uky,Lacey:2024bcm,Lacey:2025bos}.
Measurements of $v_2(p_T)$ and $v_3(p_T)$ for identified
light-flavor mesons and baryons map onto common response
functions after the contributions from initial
eccentricities, system size, medium attenuation, radial
expansion, and late-stage hadronic interactions are
scaled out. This scaling extends from the low-$p_T$
collective-flow regime to the high-$p_T$ region
increasingly sensitive to path-length-dependent partonic
energy loss, and across collision systems, centralities,
and beam energies. At fixed beam energy, the light-flavor
attenuation coefficient exhibits comparatively weak
centrality dependence after accounting for geometry
and system size, while species-dependent contributions
are scaled out separately. The resulting common response
provides a medium reference established independently
of the heavy-flavor measurements.

This independently established reference enables a direct
null test: applying the unmodified light-meson scaling
to $D^0$ anisotropies reveals whether an additional
charm-specific response is required. Any systematic
residual then identifies the scaling violation to be
quantified without retuning the common medium response.

Several features provide complementary constraints.
The centrality dependence probes changes in the
relative charm--light response where the light-flavor
attenuation coefficient varies comparatively weakly,
while the beam-energy dependence tests this difference
against changes in the common medium response.
Simultaneous $v_2$ and $v_3$ scaling provides
additional constraints through intra-harmonic and
inter-harmonic mappings. The residual low-$p_T$
response further tests whether the scaling violation
is predominantly associated with attenuation or
requires additional collective or late-stage
contributions.

The momentum scale connecting the collective-flow and
quenching regimes provides a further constraint.
In the scaling framework, the viscous correction
$\delta f=\kappa p_T^2$ is held fixed above a
data-constrained transition momentum $p_{T,\rm tr}$
\cite{Lacey:2024uky,Lacey:2024bcm,Lacey:2025bos}.
Comparison with the light-flavor reference therefore
tests whether charm differs in the transition momentum,
the strength of its anisotropic response, or both.

Small collision systems provide a complementary test.
Significant $D^0$ elliptic anisotropy has been observed
in high-multiplicity p+Pb collisions at
$\sqrt{s_{NN}}=8.16$~TeV~\cite{CMS:2018loe}.
Light-flavor scaling in p+Pb exhibits a substantially
lower transition momentum than in Pb+Pb
\cite{Lacey:2025bos}. Because the available $D^0$
measurements predominantly lie above this scale,
they test whether an additional charm--light response
difference persists in the smaller collision environment,
where the balance between collective-flow and
path-length-dependent energy-loss contributions may differ.

In this work, the established light-flavor anisotropy-response
scaling provides an independently constrained reference for
testing $D^0$ anisotropies across collision systems, centralities,
and beam energies. In Pb+Pb collisions at
$\sqrt{s_{NN}}=5.02$~TeV, identified light-flavor measurements
from ALICE are combined with CMS measurements of prompt
$D^0$ $v_2$ and $v_3$
\cite{ALICE:2018yph,CMS:2020bnz}. In ultra-central p+Pb
collisions at $\sqrt{s_{NN}}=8.16$~TeV, the $K_S^0$,
$\Lambda$, and prompt $D^0$ measurements are provided by CMS
\cite{CMS:2018loe}, while the Au+Au comparison at
$\sqrt{s_{NN}}=200$~GeV combines STAR light-flavor and
$D^0$ measurements~\cite{STAR:2015gge,STAR:2017kkh}.
The comparisons test the relative charm-response strength
and flow-to-quenching transition momentum against the
independently established light-flavor response.

%
The common light-flavor reference is anchored to the kaon
scaling function in ultra-central (uc, 0--1\%) Pb+Pb
collisions at $\sqrt{s_{NN}}=5.02$~TeV, which provides
an experimentally constrained baseline for the
medium response \cite{Lacey:2024bcm,Lacey:2024uky}.
Kaons provide a useful baseline because of their intermediate mass, 
relatively small hadronic cross section, and precise $v_{2,3}(p_T,\mathrm{cent})$
measurements. The framework has been developed and tested across
identified light-flavor mesons and baryons, collision systems,
centralities, and beam energies
\cite{Lacey:2025bos,Lacey:2024bcm,Lacey:2024fpb,Lacey:2024uky}. Only the elements
required for the present $D^0$ analysis are summarized here.

The measured anisotropies enter through the reduced response
$v_n(p_T)/\varepsilon_n$, where $\varepsilon_n$ denotes the
corresponding initial-state eccentricity. 
The scaled responses are compared as functions of
$1/\sqrt{KE_T}$, where
$KE_T=m_T-m_0=\sqrt{p_T^2+m_0^2}-m_0$
is the transverse kinetic energy and $m_0$ is the
hadron rest mass. This common kinetic-energy coordinate
facilitates comparisons across hadron species without
introducing additional scaling parameters.
The transverse size scale is represented by
$\mathbb{R}\propto\langle N_{\rm ch}\rangle^{1/3}$, while attenuation
is parameterized by $\beta=k_\beta\beta_0$. The coefficient $\beta_0$
sets the reference attenuation scale and $k_\beta$ accounts for its
beam-energy dependence. Species-dependent contributions are treated
separately within the established light-flavor scaling; for mesons,
the residual hadronic response is parameterized by
$\zeta_{\rm hs}$, with $\zeta_m=1-\zeta_{\rm hs}$.

The geometric and system-size quantities entering the scaling are
constrained independently of the anisotropy-response transformation.
The eccentricities $\varepsilon_2$ and $\varepsilon_3$ are obtained
from Monte Carlo Glauber calculations of the initial collision
geometry for the corresponding collision system and event selection
\cite{Alver:2008zza,Lacey:2013eia,
Lacey:2024bcm,Lacey:2024uky}. The charged-particle multiplicities are
taken from measurements appropriate to each collision system and
event class. For Pb+Pb collisions at
$\sqrt{s_{NN}}=5.02$~TeV, the corresponding
$\langle N_{\rm ch}\rangle$ values are obtained from the ALICE
charged-particle multiplicity measurements
\cite{ALICE:2015juo}; the p+Pb and Au+Au analyses likewise use the
measured multiplicities appropriate to their respective event
selections, expressed within the common pseudorapidity interval used
for the scaling. The dimensionless transverse-size variable
$\mathbb R$ is then constructed from
$\langle N_{\rm ch}\rangle^{1/3}$ using the same prescription as in
the established light-flavor scaling analyses
\cite{Lacey:2024bcm,Lacey:2024fpb,Lacey:2024uky}. 
Consequently, the eccentricities and measured
charged-particle multiplicities determine the
geometric and transverse-size inputs to the scaling,
without adjustment to obtain data collapse.
 When measurements with different experimental centrality
or multiplicity selections are compared, each is transformed using
the geometric and multiplicity inputs appropriate to its respective
event class. 
The momentum-dependent viscous correction entering the
scaling relations is parameterized as
$\delta f(p_T)=\kappa\min(p_T^2,p_{T,\rm tr}^2)$.
Consistent with this transition prescription, the
species-dependent corrections $\zeta_{\rm rf}$ and
$\zeta_{\rm hs}$ are applied below their respective
$p_{T,\rm tr}$ and set to zero above it.
For comparisons across collision environments, the
fixed Pb+Pb reference uses $\delta f_{\rm ref}(p_T)$,
while the comparison system uses $\delta f_X(p_T)$,
with their respective transition momenta and the
same independently constrained coefficient $\kappa$.

The meson $v_2$ scaling relation is
\begin{multline}
\frac{v_2(p_T,\mathrm{uc})}{\varepsilon_2(\mathrm{uc})}\,
e^{\tfrac{2\beta_0}{\mathbb{R}_{\rm uc}}
(2+\delta f_{\rm ref}(p_T))}
=
e^{\alpha\,\tfrac{2\beta_0}{\mathbb{R}_{\rm uc}}
(2+\delta f_{\rm ref}(p_T))\,\zeta_M^{(2)}}
\\[-2pt]
\times
\left(\frac{v_2'(p_T)}{\varepsilon_2'}\right)^{\zeta_m}\,
e^{\tfrac{2\zeta_m\beta}{\mathbb{R}_{\rm uc}}
\left(\tfrac{\mathbb{R}_{\rm uc}}{\mathbb{R}'}-1\right)
(2+\delta f_X(p_T))},
\label{eq:v2_scaling_mesons}
\end{multline}
where primes denote the comparison system or centrality and
$\zeta_M^{(2)}=\zeta_m+\gamma_{32}^{X}+(1-k_\beta)$.
The geometry-only offset
$\gamma_{32}^{X}\equiv
\ln[(\varepsilon_3/\varepsilon_2)_{\rm ref}/
(\varepsilon_3/\varepsilon_2)_{\rm sys}]|_X$
is fixed by the eccentricity ratios, with
$\gamma_{32}^{X}=0$ for all Pb+Pb centralities.
For p+Pb, its sign is reversed to account for the
inverted centrality evolution of $\varepsilon_3/\varepsilon_2$
relative to the large-$A+A$ reference.
This offset enters only through the normalization exponents
and is not independently adjusted for charm.

In applying this scaling relation, the left-hand side remains
the fixed ultra-central reference, with all centrality-dependent
modifications implemented on the right-hand side.
The normalization factor is controlled by
$\alpha=0$, $0.5$, and $1$ for non-ultra-central,
near-ultra-central, and ultra-central selections,
respectively. Thus, the normalization exponential
is unity for non-ultra-central collisions and retains
its prescribed $p_T$ dependence for ultra-central selections,
with near-ultra-central selections providing an
intermediate response. The comparison-system attenuation
exponential retains its prescribed
$(2+\delta f_X(p_T))$ dependence for all selections.

An independent constraint is provided by the inter-harmonic mapping
$v_3/\varepsilon_3\!\to\!v_2/\varepsilon_2$,
\begin{multline}
\frac{v_2(p_T,\mathrm{uc})}{\varepsilon_2(\mathrm{uc})}\,
e^{\tfrac{2\beta_0}{\mathbb{R}_{\rm uc}}
(2+\delta f_{\rm ref}(p_T))}
=
e^{\tfrac{2\beta_0}{\mathbb{R}_{\rm uc}}\,
\zeta_M^{(2\leftarrow3)}}
\\[-2pt]
\times
\left(\frac{v_3'(p_T)}{\varepsilon_3'}\right)^{
\tfrac{2}{3}\zeta_m}\,
e^{\tfrac{2\zeta_m\beta}{\mathbb{R}_{\rm uc}}
\left(\tfrac{\mathbb{R}_{\rm uc}}{\mathbb{R}'}
-1\right)(3+\delta f_X(p_T))},
\label{eq:v2_from_v3_mesons}
\end{multline}
where
$\zeta_M^{(2\leftarrow3)}=1-\zeta_M^{(2)}$.
The complementary $2\!\to\!2$ and $3\!\to\!2$ relations therefore
provide independent constraints on the mapping to the same reference
response. 
\begin{figure*}[t]
\centering
\includegraphics[width=0.75\textwidth]
{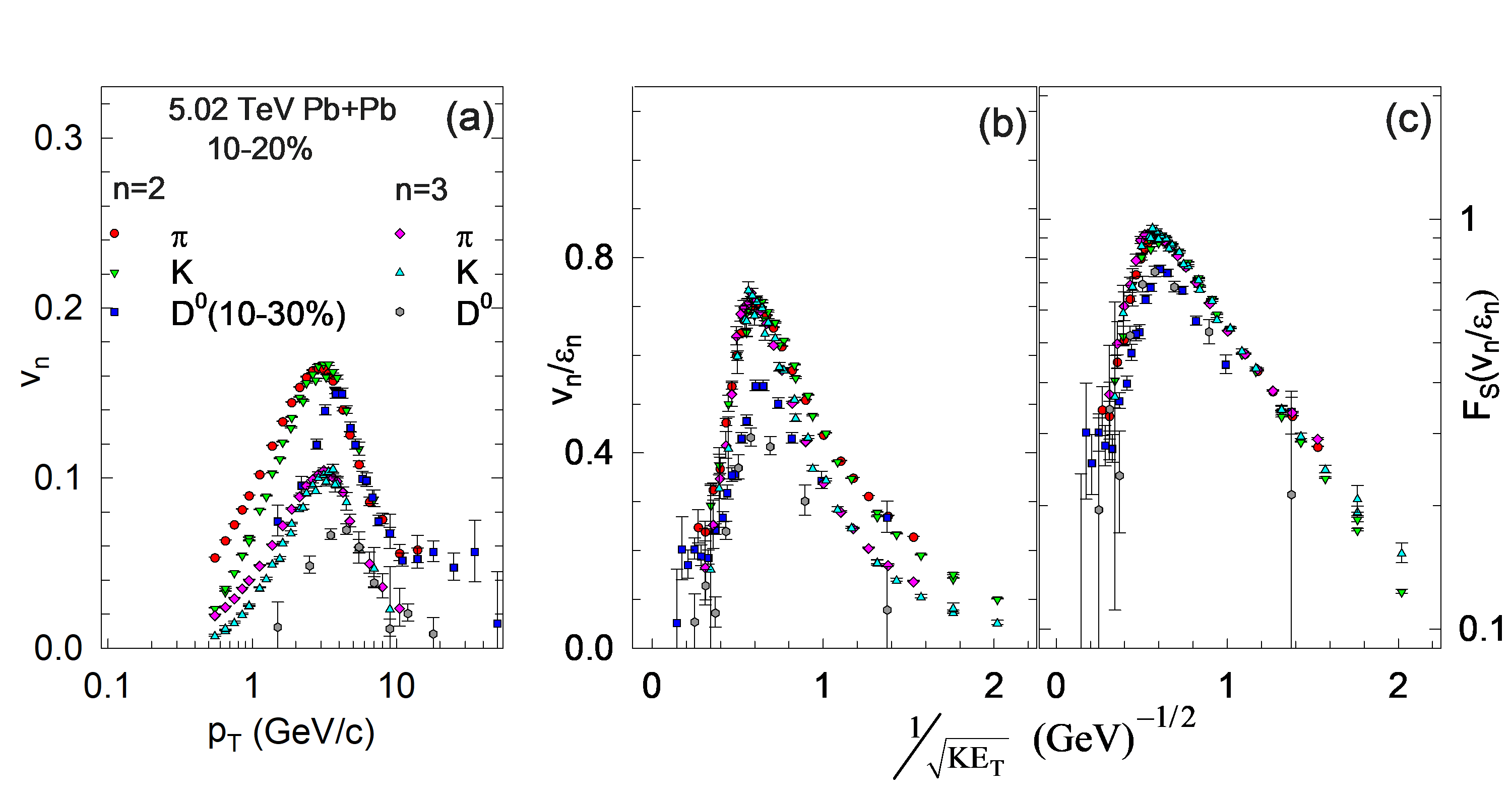}
\caption{
(Color online) Anisotropy-response scaling for $\pi$, $K$, and
$D^0$ mesons in Pb+Pb collisions at
$\sqrt{s_{NN}}=5.02$~TeV using the established light-meson
prescription without charm-specific modification.
The $\pi$ and $K$ measurements correspond to the ALICE 10--20\%
centrality interval \cite{ALICE:2018yph} and the $D^0$ measurements to the CMS 10--30\%
interval \cite{CMS:2020bnz}; each is scaled using the inputs
appropriate to its respective centrality selection.
Panels (a)--(c) show the measured anisotropies, the
eccentricity-scaled responses, and the fully scaled responses,
respectively. The transformed $D^0$ harmonics exhibit approximate
inter-harmonic closure but remain displaced from
the light-meson reference.
}
\label{fig:D0_mesonscaling}
\end{figure*}

The piecewise definition of $\delta f(p_T)$ ensures
continuous matching of the scaling correction at
$p_{T,\rm tr}$, which characterizes the effective
transition between collective-flow-dominated and
quenching-sensitive anisotropic responses.
In each collision environment, the correction grows
as $\kappa p_T^2$ below its respective $p_{T,\rm tr}$
and remains fixed at $\kappa p_{T,\rm tr}^2$ above it,
without constraining the intrinsic momentum dependence
of the high-$p_T$ quenching response.
The variation of this transition scale across collision
environments provides an additional constraint on
parton--medium dynamics, distinct from the attenuation
strength. A lower $p_{T,\rm tr}$ corresponds to an earlier
crossover and a smaller matching correction for
fixed $\kappa$. For the light-flavor Pb+Pb reference at
$\sqrt{s_{NN}}=5.02$~TeV,
$p_{T,\rm tr,ref}\simeq4.5$--$5.0$~GeV/$c$
\cite{Lacey:2024fpb,Lacey:2024uky}.

The $D^0$ scaling test begins by treating the $D^0$ as a meson
within this established framework. The geometric, system-size,
and medium-response inputs are constrained by the light-flavor
scaling and applied without charm-specific modification, using
the appropriate centrality-dependent inputs for each event class.
The attenuation structure and normalization are likewise retained
unchanged. The meson response follows the established prescription
$\zeta_m=1-\zeta_{\rm hs}$, with $\zeta_{\rm hs}$ retaining
its definition and role in the light-flavor scaling framework.
The same prescription is applied to the $2\!\to\!2$ and,
where available, $3\!\to\!2$ mappings, providing a direct
null test of whether the measured $D^0$ anisotropies follow
the independently established light-flavor response.

A systematic departure of the transformed $D^0$ measurements
from the common reference defines the $D^0$ scaling violation.
No charm-specific response coefficient is introduced in
establishing this violation, nor are the underlying medium-response
parameters independently retuned for charm. Any additional
response is introduced only after the violation is identified,
using a minimal extension of the established meson transformation
to recover simultaneous scaling of the available harmonics.
The fidelity of the transformed responses against the
independently established common scaling function
therefore provides a differential constraint on the
additional charm response.

%
The null test is illustrated in Fig.~\ref{fig:D0_mesonscaling}
for Pb+Pb collisions at $\sqrt{s_{NN}}=5.02$~TeV.
For this comparison, the light-meson measurements are provided
by ALICE $\pi$ and $K$ data in the 10--20\% centrality
interval~\cite{ALICE:2018yph}, while the higher-statistics
CMS $D^0$ $v_2$ and $v_3$ measurements correspond to
10--30\%~\cite{CMS:2020bnz}.
Both datasets employ the scalar-product method,
minimizing differences associated with the anisotropy
estimator. The light-meson and $D^0$ measurements need
not have identical centrality selections, as each dataset
is transformed using the eccentricities and the
multiplicity-derived transverse-size scale appropriate
to its respective event selection.
The scaling therefore compares the responses after these
geometric and system-size contributions have been scaled out,
rather than treating the two centrality intervals as equivalent.
\begin{figure*}[t]
\centering
\includegraphics[width=0.75\textwidth]
{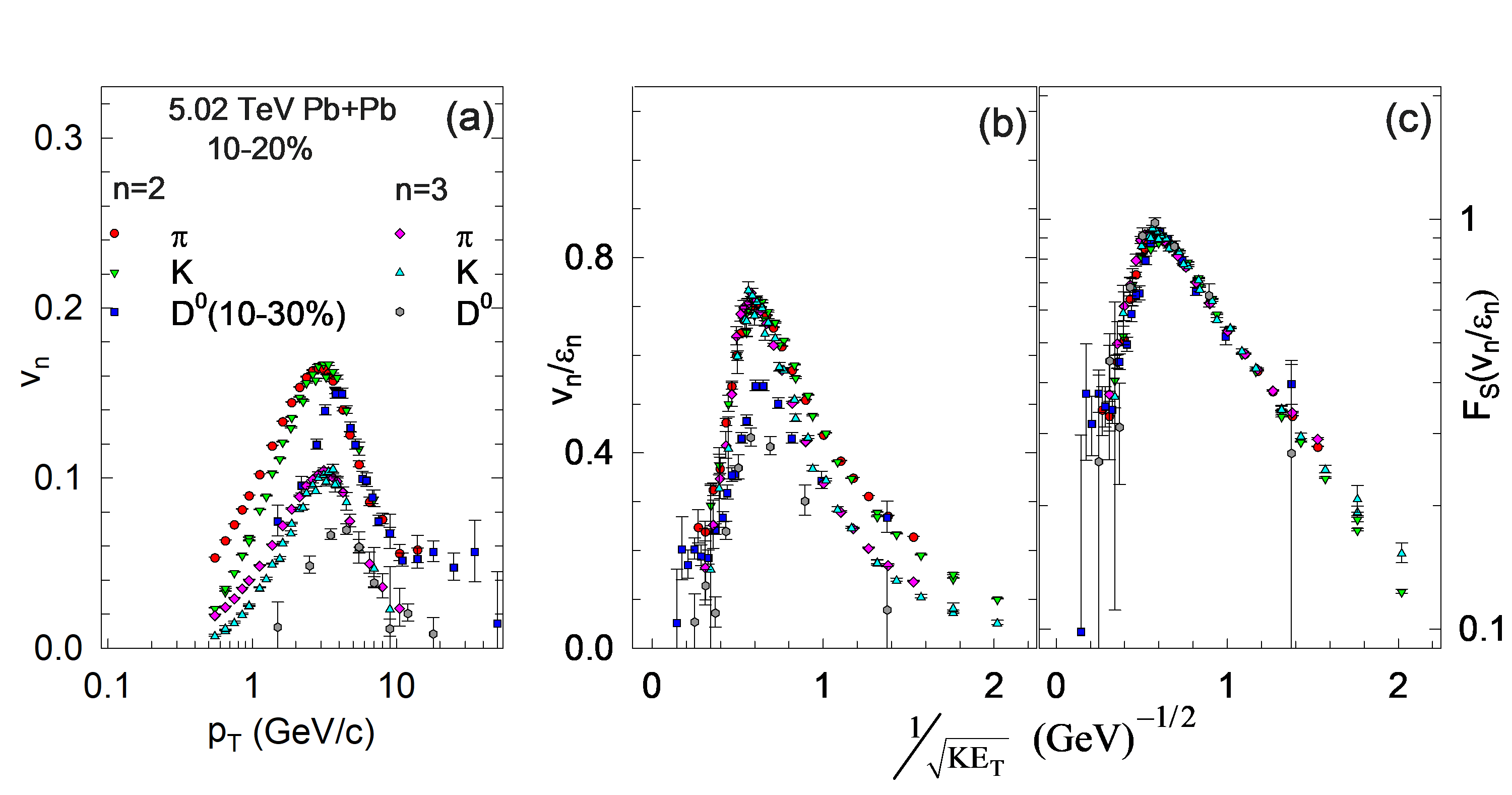}
\caption{
(Color online) Same measurements and centrality selections as
Fig.~\ref{fig:D0_mesonscaling}, with the additional $D^0$ response
implemented through $\beta_0\rightarrow k_c\beta_0$,
with $k_c\simeq2.0$, and $\zeta_{\rm hs}\simeq0.06$.
The light-meson scaling prescription remains unchanged.
Panels (a)--(c) show the measured anisotropies, the
eccentricity-scaled responses, and the fully scaled responses,
respectively. The modified $D^0$ response recovers the
common light-meson scaling function.
}
\label{fig:D0_charmscaling}
\end{figure*}

\begin{figure*}[t]
\centering
\includegraphics[width=0.75\textwidth]
{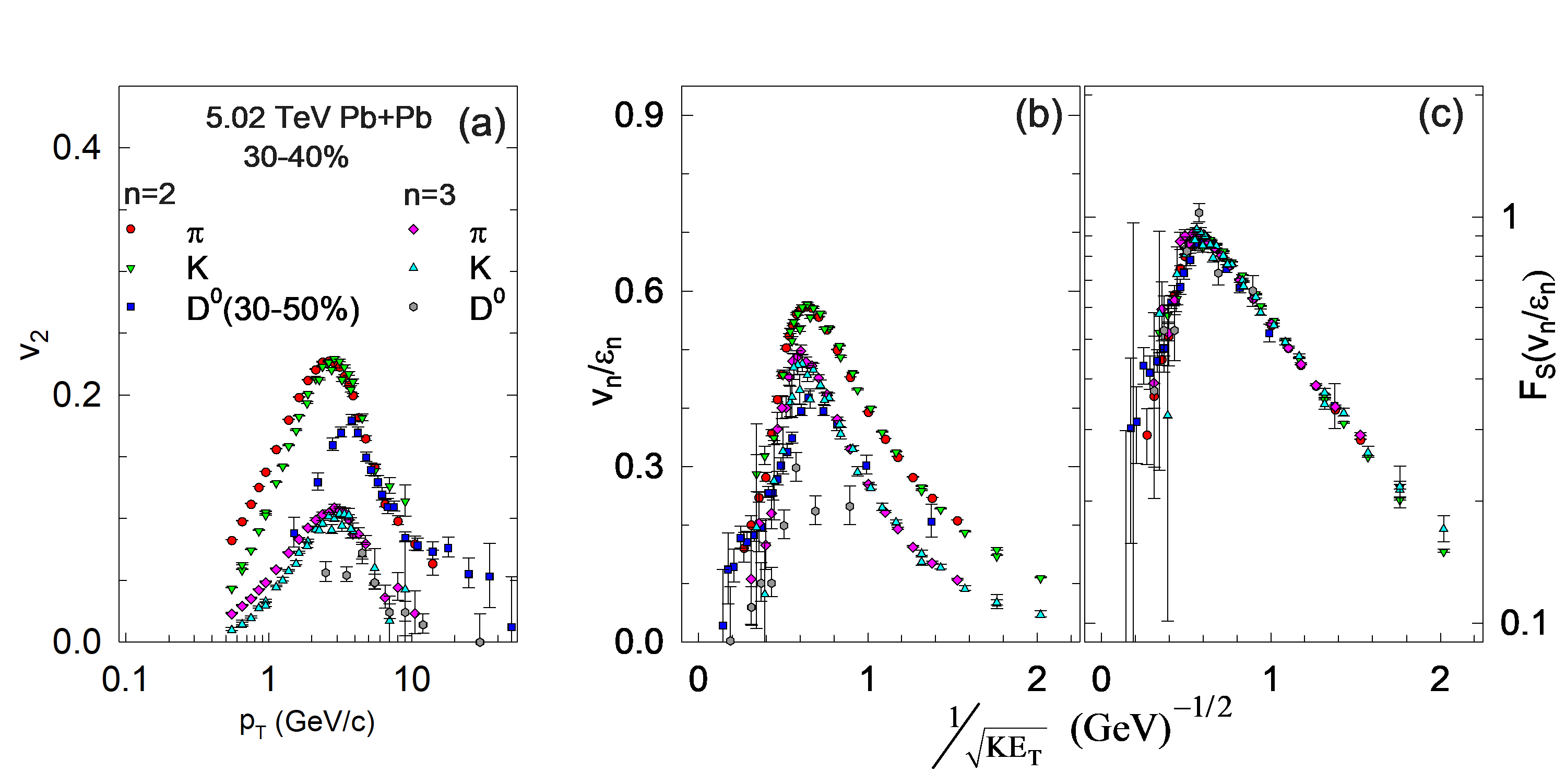}
\caption{
(Color online) Anisotropy-response scaling for $\pi$, $K$, and
$D^0$ mesons in more peripheral Pb+Pb collisions at
$\sqrt{s_{NN}}=5.02$~TeV. The $\pi$ and $K$ measurements correspond
to the 30--40\% centrality interval and the $D^0$ measurements to
30--50\%; each is scaled using the inputs appropriate to its
respective centrality selection. The $D^0$ response is transformed
with $\zeta_{\rm hs}\simeq0.06$ and $k_c\simeq1.27$, while the
light-meson prescription remains unchanged. Panels (a)--(c) show the
measured anisotropies, the eccentricity-scaled responses, and the
fully scaled responses, respectively. 
}
\label{fig:D0_peripheral}
\end{figure*}

For context, the light-flavor scaling coefficients are independently
constrained by the identified-hadron measurements
\cite{Lacey:2024uky,Lacey:2024bcm,Lacey:2025bos}. 
In Pb+Pb collisions at $\sqrt{s_{NN}}=5.02$~TeV, the 
common attenuation response is characterized by $k_\beta\simeq1$, 
with $\zeta_{\rm hs}\simeq0$ for light mesons. The light-meson
transition momentum is likewise constrained independently
of the $D^0$ measurements. The baryon sector maps onto the
same common scaling function after its stronger radial-flow
response is scaled out, with $\zeta_{\rm rf}$ increasing
approximately logarithmically with $\langle N_{\rm ch}\rangle$
and reaching approximately $0.46$ in central collisions.
Thus, substantial species- and centrality-dependent effects,
including the stronger mass-dependent collective response
of baryons, are already accommodated within the light-flavor
framework. The $D^0$ comparison therefore tests for an
additional charm--light response difference relative to this
independently constrained medium-response reference.

Figure~\ref{fig:D0_mesonscaling}(a) shows the measured $v_2$ and $v_3$
before the scaling transformation. The species and harmonic
dependences are clearly visible in the measured anisotropies. After
scaling by the corresponding initial eccentricities,
Fig.~\ref{fig:D0_mesonscaling}(b), the pion and kaon $v_2$ and $v_3$
responses exhibit closely related trends, while the $D^0$ responses
remain systematically separated from the light-meson sector.
The fully transformed responses, denoted $F_S(v_n/\varepsilon_n)$
in Fig.~\ref{fig:D0_mesonscaling}(c), bring both harmonics for pions
and kaons onto the common light-meson scaling function, demonstrating
simultaneous $2\!\to\!2$ and $3\!\to\!2$ closure in the light sector.
Notably, the transformed $D^0$ $v_2$ and $v_3$ measurements also
approximately collapse onto a common response, but this response is
systematically displaced below the light-meson scaling function.
Thus, the $D^0$ scaling violation is not primarily a breakdown of the
inter-harmonic mapping. Instead, the two harmonics exhibit a coherent
displacement from the independently constrained light-meson
reference, identifying a distinct charm--light response difference
after the established common contributions are scaled out.

The result in Fig.~\ref{fig:D0_mesonscaling} therefore establishes
the null-test failure before any charm-specific modification is
introduced. The coherence of the transformed $D^0$ $v_2$ and $v_3$
responses further indicates that the required modification need not
independently alter the two harmonics. The next step is therefore to 
identify the minimal extension of the established meson transformation
needed to map the displaced $D^0$ response onto
the common light-meson scaling function.
 The attenuation response is tested by rescaling
the reference attenuation strength as
$\beta_0\rightarrow k_c\beta_0$, where $k_c=1$ corresponds
to the unchanged light-meson attenuation response.
Importantly, $k_c$ enters only through the $D^0$ attenuation
terms; $k_\beta$ and the associated normalization exponents
are retained unchanged from the light-flavor reference.
The meson-response factor $\zeta_m=1-\zeta_{\rm hs}$
is retained with its established definition and
role in the light-flavor scaling framework.

Figure~\ref{fig:D0_charmscaling} shows the result of applying this
minimal extension to the same measurements. The pion and kaon $v_2$
and $v_3$ measurements remain on the common light-meson scaling
function because their prescription is unchanged.
The established scaling fidelity across harmonics, centralities,
particle species, collision systems, and beam energies provides
multiple complementary constraints on this common response.
For the $D^0$, the scaling violation is resolved by rescaling
the reference attenuation strength,
$\beta_0\rightarrow k_c\beta_0$, with $k_c\simeq2.0$,
together with a small hadronic-rescattering response
coefficient, $\zeta_{\rm hs}\simeq0.06$.
The transformed $D^0$ $v_2$ and $v_3$ responses
thereby preserve their inter-harmonic consistency
while recovering the common scaling function.
The modest low-$p_T$ contribution from
$\zeta_{\rm hs}$ is distinguishable from the dominant
$k_c$ modification: varying $\zeta_{\rm hs}$
degrades simultaneous $D^0$ $v_2$ and $v_3$
scaling below the transition, even after
readjusting $k_c$.

The scaling provides distinct sensitivity to $k_c$ and
$p_{T,\rm tr}$. The coefficient $k_c$ rescales the reference
attenuation strength entering the $D^0$ transformation over the full
momentum range and therefore controls the overall mapping onto the
light-meson scaling function. Its value is identified through the
simultaneous alignment of the available $D^0$ harmonics with the
independently established light-meson reference, rather than by
fitting the reference response to the charm data.
Below $p_{T,\rm tr}$, simultaneous $D^0$ $v_2$
and $v_3$ scaling provides complementary sensitivity
to $\zeta_{\rm hs}$ and $k_c$, while above the
transition the additional charm response is governed
by the attenuation rescaling $k_c$.
In contrast, $p_{T,\rm tr}$ controls the momentum dependence
of the matching prescription, with inconsistent transition
scales producing systematic deviations at intermediate and
high $p_T$. Thus, $k_c$ and $p_{T,\rm tr}$ constrain distinct
features of the $D^0$ response rather than providing
interchangeable rescalings.

The comparison of Figs.~\ref{fig:D0_mesonscaling} and
\ref{fig:D0_charmscaling} highlights the physical
significance of the charm--light response difference.
The kaon-anchored scaling function provides an
independently constrained mesonic reference within
the same collision environment. The coefficient $k_c\simeq2.0$ 
indicates substantially weaker development of the $D^0$ anisotropic 
response relative to the light-meson reference, while the
smaller $\zeta_{\rm hs}\simeq0.06$ retains its
established role in the hadronic-rescattering response channel.

At low and intermediate $p_T$, this is consistent
with weaker development of the anisotropic
collective-flow response, while at high $p_T$
it is consistent with a reduced quenching-sensitive
anisotropic response relative to the light-meson
reference. Thus, $k_c$ provides a
differential constraint on the charm response
to the same evolving medium.

The momentum dependence provides an additional
constraint. The $D^0$ measurements are consistent
with a transition momentum of approximately
$4.5$--$5.0$~GeV/$c$, comparable within the present
resolution to the light-flavor value despite the
additional charm attenuation characterized by
$k_c\simeq2.0$. The charm--light difference is
therefore expressed primarily through the strength
of the anisotropic response rather than through
an appreciable shift in the transition momentum.
This distinction motivates a further test of the
centrality dependence of the relative charm response.

The centrality dependence of the relative charm--light response is
examined in Fig.~\ref{fig:D0_peripheral}, where the same scaling
procedure is applied to the more peripheral Pb+Pb measurements. 
For this comparison, the light-meson measurements
are provided by $\pi$ and $K$ data in the 30--40\%
interval, while the $D^0$ measurements correspond to 30--50\%.
 As in the preceding comparison, each dataset is transformed
using the geometric and system-size inputs appropriate
to its respective centrality selection. Figures~\ref{fig:D0_peripheral}(a) and
\ref{fig:D0_peripheral}(b) show that the measured and
eccentricity-scaled $D^0$ responses remain distinct from the
light-meson sector. After the full scaling transformation,
Fig.~\ref{fig:D0_peripheral}(c), the $D^0$ $v_2$ and $v_3$ responses
collapse onto the kaon-anchored light-meson scaling function. In
contrast to the $k_c\simeq2.0$ required for the more central
collisions, the 30--50\% $D^0$ measurements recover the common
response with $k_c\simeq1.27$.
\begin{figure*}[t]
\centering
\includegraphics[width=0.75\textwidth]
{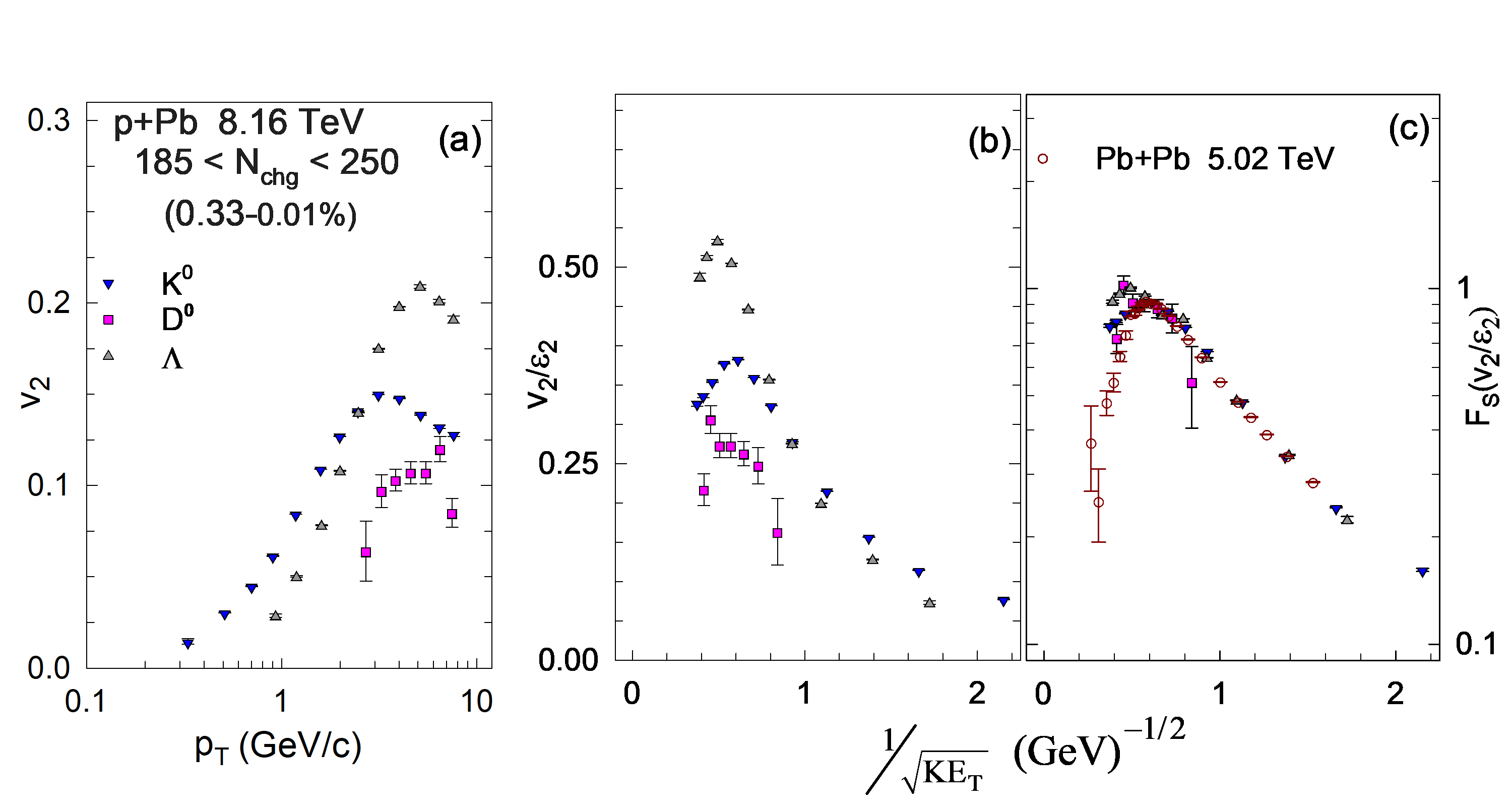}
\caption{
(Color online) Anisotropy-response scaling for $K^0_S$, $\Lambda$,
and $D^0$ hadrons in ultra-central ($0.33$--$0.01$\%) p+Pb collisions
at $\sqrt{s_{NN}}=8.16$~TeV. Panels (a)--(c) show the measured $v_2$,
the eccentricity-scaled responses, and the fully scaled responses, respectively.
The $D^0$ scaling uses $k_c\simeq1.5$; the Pb+Pb reference
response at $\sqrt{s_{NN}}=5.02$~TeV is shown in panel (c).
}
\label{fig:D0_pPb}
\end{figure*}

The decrease in $k_c$ establishes a centrality dependence
of the relative charm--light response. 
A separate scaling check of the 0--10\% interval
also yields $k_c\simeq2.0$, indicating that the
larger charm--light response difference persists
from central through mid-central Pb+Pb collisions
before decreasing to $k_c\simeq1.27$ in the
30--50\% interval. This evolution contrasts
with the comparatively weak centrality dependence
of $k_\beta$ for light-flavor hadrons at the same
beam energy. Thus, the narrowing charm--light
response difference toward peripheral collisions
provides a constraint on the evolution of heavy-quark
dynamics with the changing medium, after the
established geometry, system-size, and light-flavor
response dependencies have been scaled out.

%
A complementary test of the charm response is provided by
ultra-central p+Pb collisions at $\sqrt{s_{NN}}=8.16$~TeV. The
$0.33$--$0.01\%$ event class, corresponding to the reported
multiplicity selection $185<N_{\rm chg}<250$, provides a compact
collision system with a substantially different medium extent from
that of Pb+Pb collisions. For consistency with the system-size
definition used throughout the scaling analysis, the reported
multiplicity is converted to the corresponding
$\langle N_{\rm ch}\rangle$ within $|\eta|<0.5$ before constructing
$\mathbb R\propto\langle N_{\rm ch}\rangle^{1/3}$. The eccentricity
and multiplicity inputs appropriate to this p+Pb event class are
therefore incorporated independently of the anisotropy-response
transformation.

Figure~\ref{fig:D0_pPb} applies the same anisotropy-response scaling
framework to $K^0_S$, $\Lambda$, and $D^0$ measurements in these
ultra-central p+Pb collisions. The inclusion of the $\Lambda$
provides a baryon reference and tests whether the $D^0$ behavior can
be accounted for by the mass- and species-dependent response already
incorporated in the light-flavor scaling framework. The measured
$v_2$ values in Fig.~\ref{fig:D0_pPb}(a) exhibit substantial species
dependence, which remains evident after eccentricity scaling in
Fig.~\ref{fig:D0_pPb}(b). After the full scaling transformation,
Fig.~\ref{fig:D0_pPb}(c), the low- and intermediate-$p_T$ $K^0_S$
and $\Lambda$ responses map onto the common Pb+Pb-anchored
light-flavor scaling function when the lower p+Pb transition
momenta appropriate to the meson and baryon sectors are used.
At higher $p_T$, residual light-flavor scaling deviations
may reflect weaker quenching-sensitive anisotropy, with
jet-driven nonflow correlations becoming relatively more
prominent in the smaller p+Pb system.
The $D^0$ measurements recover the common response
with $k_c\simeq1.5$. Since these measurements lie
above the transition momentum, they do not
independently constrain $\zeta_{\rm hs}$.

The common low- and intermediate-$p_T$ scaling of the $K^0_S$ and
$\Lambda$, together with the transformed $D^0$ response, is
particularly informative for interpreting the charm modification.
The light meson and baryon reach the same reference response after
their established species-dependent contributions are scaled out,
including the stronger radial-flow response of the baryon sector and
the lower transition momenta appropriate to the p+Pb system. The
additional modification required for the $D^0$ therefore cannot be
attributed simply to its large mass or to the meson--baryon response
difference. Instead, $k_c\simeq1.5$ demonstrates that a measurable
charm--light response difference persists even in the
compact medium produced in ultra-central p+Pb collisions.
\begin{figure*}[t]
\centering
\includegraphics[width=0.75\textwidth]
{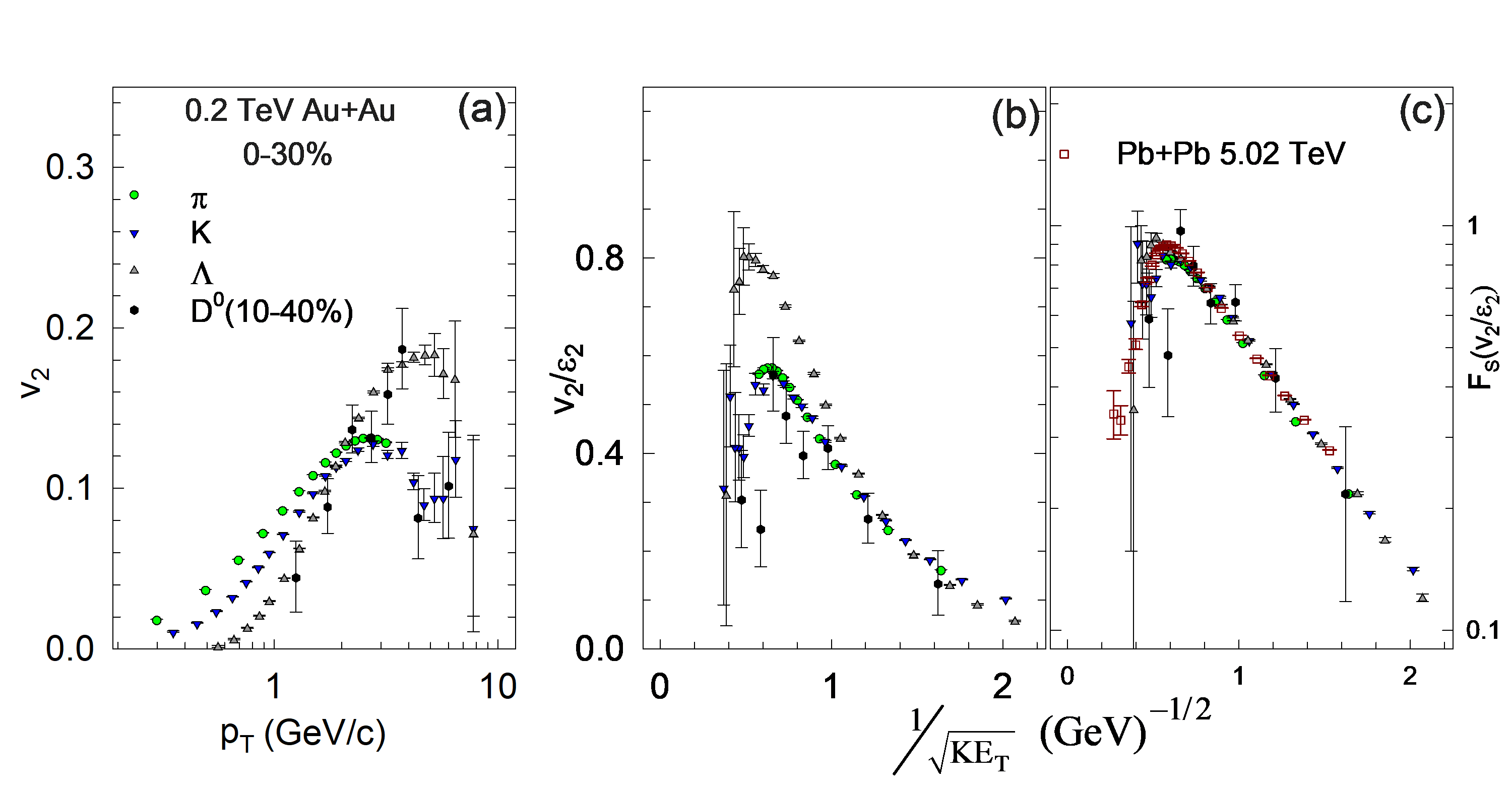}
\caption{
(Color online) Anisotropy-response scaling for $\pi$, $K$,
$\Lambda$, and $D^0$ hadrons in Au+Au collisions at
$\sqrt{s_{NN}}=200$~GeV. The light-flavor measurements correspond
to 0--30\% centrality and the $D^0$ measurements to 10--40\%;
each is scaled using the inputs appropriate to its
respective centrality selection.
Panels (a)--(c) show the measured $v_2$, the eccentricity-scaled
response, and the fully scaled response, respectively.
The transformed $D^0$ response is consistent with
the common light-flavor scaling function without
additional charm attenuation rescaling
($k_c\simeq1$).
}
\label{fig:D0_RHIC}
\end{figure*}

The momentum dependence of the p+Pb scaling provides an additional
constraint. For light mesons, recovery of the common response at low
and intermediate $p_T$ requires a transition momentum near
$p_{T,\rm tr}\simeq1.6$~GeV/$c$, substantially below the
approximately $4.5$--$5.0$~GeV/$c$ scale in Pb+Pb collisions; the
corresponding baryon transition occurs at a higher momentum, near
$3$~GeV/$c$. Within the matching prescription, these lower transition
momenta identify an earlier crossover from the viscously attenuated
collective-flow response to the regime increasingly sensitive to
path-length-dependent partonic energy loss.
Above the transition, the scaling correction remains
fixed at $\kappa p_{T,\rm tr}^2$. Since the available
$D^0$ measurements begin above approximately
$2$~GeV/$c$, they test consistency with the lower
transition scale independently established by light
mesons, rather than directly locating a distinct
charm crossover. Thus, as in Pb+Pb collisions, the charm--light
difference is expressed primarily through the strength of the
anisotropic response rather than through an observable displacement
of the transition scale.

The value $k_c\simeq1.5$ provides an important counterpoint to the
Pb+Pb centrality dependence established earlier. In Pb+Pb collisions at
$\sqrt{s_{NN}}=5.02$~TeV, $k_c$ decreases from approximately $2.0$
in central and mid-central events to approximately $1.27$ in the
30--50\% interval, indicating that the charm--light response
difference narrows as the properties and space-time evolution of the
Pb+Pb medium change. The ultra-central p+Pb result does not simply
continue this evolution toward the light-flavor limit, $k_c=1$,
despite the substantially smaller collision system. Geometric extent
alone therefore does not determine the relative charm response. The
p+Pb measurement is obtained at the higher beam energy of
$\sqrt{s_{NN}}=8.16$~TeV and in a collision environment with
different geometry, density, lifetime, and relative contributions
from collective flow and path-length-dependent partonic energy loss.
The observed $k_c\simeq1.5$ therefore indicates that the
charm--light anisotropic-response difference depends more broadly on
the properties and space-time evolution of the produced medium,
rather than on system size alone.

%
The large beam-energy lever arm between RHIC and the LHC provides an
additional test of the relative charm--light response.
Figure~\ref{fig:D0_RHIC} compares $\pi$, $K$, $\Lambda$, and $D^0$
measurements in Au+Au collisions at $\sqrt{s_{NN}}=200$~GeV. The
light-flavor measurements correspond to the 0--30\% centrality
interval, while the $D^0$ measurements correspond to 10--40\%.
The STAR light-flavor and $D^0$ $v_2$ measurements employ
comparable $\eta$-subevent event-plane methods, minimizing
differences associated with the anisotropy estimator.
As in the preceding comparisons, each dataset is transformed using
the geometric and system-size inputs appropriate to its respective
event selection.

The measured $v_2$ values in Fig.~\ref{fig:D0_RHIC}(a) exhibit a
pronounced species dependence, which remains evident after
eccentricity scaling in Fig.~\ref{fig:D0_RHIC}(b). Application of the
full scaling transformation, Fig.~\ref{fig:D0_RHIC}(c), maps the
light mesons and the $\Lambda$ onto the common light-flavor response
after their established species-dependent contributions are scaled
out. Within the precision of the available measurements, the $D^0$
$v_2$ is also consistent with this common response without an
additional charm attenuation rescaling, corresponding to
$k_c\simeq1$.

The simultaneous comparison with the $\Lambda$ is particularly useful
for interpreting this result. Despite its substantially larger mass
and the strong species dependence visible before the full
transformation, the $\Lambda$ maps onto the same light-flavor scaling
function after its established baryon response is scaled out. The
consistency of the $D^0$ with this response at $k_c\simeq1$ therefore
cannot be attributed simply to its large mass. Rather, it indicates
that at $\sqrt{s_{NN}}=200$~GeV the measured charm anisotropy requires
little additional attenuation rescaling relative to the independently
constrained light-flavor medium response.

This result is qualitatively different from that obtained for central
and mid-central Pb+Pb collisions at the LHC. The light-flavor
attenuation response at $\sqrt{s_{NN}}=200$~GeV is independently
characterized by $k_\beta\simeq0.63$, compared with
$k_\beta\simeq1$ for the 5.02-TeV Pb+Pb reference. Thus,
$k_c\simeq1$ at RHIC does not imply that the underlying attenuation
response or medium conditions are the same at the two beam energies.
Rather, it indicates that little additional charm attenuation
rescaling is required relative to the light-flavor response
appropriate to the RHIC medium. At the LHC, by contrast, the Pb+Pb
measurements require $k_c>1$, reaching approximately $2$ in central
and mid-central collisions. The RHIC--LHC comparison therefore reveals 
differences in both the common light-flavor attenuation response,
characterized by $k_\beta$, and the additional
charm--light response, characterized by $k_c$.
Together, these differences provide a differential
constraint on heavy-quark dynamics across collision
environments with distinct temperatures, densities,
geometries, and space-time evolution.

The RHIC constraint is less stringent than its LHC counterpart because
the available $D^0$ measurement provides $v_2$ without the
simultaneous $v_3$ information needed for the inter-harmonic
$3\!\to\!2$ closure test. Nevertheless, the consistency of the
$D^0$ $v_2$ with the independently established 200-GeV light-flavor
response provides an important constraint on the beam-energy
dependence of the relative charm response. Together, the Pb+Pb,
p+Pb, and Au+Au measurements demonstrate that $k_c$ is not a fixed
property of the $D^0$, but quantifies an additional charm--light
response difference that evolves with the collision environment.

%
The central result is that the independently constrained
light-flavor scaling provides a direct null test of the
charm anisotropic response. In Pb+Pb collisions at
$\sqrt{s_{NN}}=5.02$~TeV, the transformed $D^0$ $v_2$
and $v_3$ exhibit approximate inter-harmonic closure
but remain systematically displaced from the common
light-meson scaling function. 
This violation is resolved primarily through the
charm-specific attenuation rescaling $k_c$, together
with a small low-$p_T$ contribution through the
established hadronic-rescattering response channel,
$\zeta_{\rm hs}\simeq0.06$, without modifying the
independently constrained light-flavor reference.
The pronounced centrality, collision-system, and
beam-energy dependence of $k_c$ demonstrates that the
relative charm response evolves with the collision
environment.

The physical interpretation of $k_c$ follows from its role in
the scaling transformation. It rescales the reference
attenuation strength only in the $D^0$ attenuation terms,
$\beta_0\rightarrow k_c\beta_0$, without modifying the
independently constrained geometry, system size,
light-flavor coefficient $k_\beta$, or normalization
structure. Values $k_c>1$ indicate weaker development
of the $D^0$ anisotropic response relative to the
light-meson reference. At low and intermediate $p_T$,
this difference is reflected in the collective-flow
response, while at high $p_T$ it is reflected in the
quenching-sensitive anisotropic response. 
Thus, $k_c$ provides a differential constraint on
heavy-quark transport across the collective-flow and
quenching-sensitive momentum regimes, relative to
the independently established light-flavor response.

The kaon-anchored reference provides an important control
on this interpretation. The established light-flavor scaling
demonstrates that mesons and baryons, despite their different
masses and species-dependent responses, map onto the same
common function once the corresponding contributions are
scaled out. This behavior is established in Pb+Pb collisions
and further illustrated by the $\Lambda$ comparisons in
p+Pb and Au+Au. The additional $D^0$ modification
therefore cannot be attributed simply to conventional
mass ordering or meson--baryon response differences.
Moreover, the modest correction associated with
$\zeta_{\rm hs}\simeq0.06$ reinforces the conclusion
that the principal charm--light difference is
captured by the attenuation rescaling $k_c$.

The momentum dependence provides a complementary
constraint. In Pb+Pb collisions at
$\sqrt{s_{NN}}=5.02$~TeV, the $D^0$ and light-meson
responses are consistent with comparable transition
momenta, $p_{T,\rm tr}\simeq4.5$--$5.0$~GeV/$c$,
despite the substantial charm--light attenuation
difference characterized by $k_c\simeq2.0$.
The transition connects collective-flow-dominated
and quenching-sensitive anisotropic responses,
whose underlying momentum-transport dynamics
are theoretically related~\cite{Majumder:2007zh}.
The comparable transition scales suggest that
the effective crossover is strongly influenced
by the common collision environment, whereas
the attenuation response retains a pronounced
flavor dependence. Thus, $k_c$ and $p_{T,\rm tr}$
provide complementary constraints on heavy-quark
transport and its coupling to the evolving medium.

The ultra-central p+Pb results further demonstrate
that the relative charm response is not governed
by system size alone: $k_c\simeq1.5$ does not
follow a simple monotonic extrapolation of the
Pb+Pb centrality trend. The light-meson transition
scale, $p_{T,\rm tr}\simeq1.6$~GeV/$c$, is
substantially lower than in Pb+Pb, and the available
$D^0$ measurements remain consistent with this
scale. At higher $p_T$, residual light-flavor
scaling deviations may reflect weaker
path-length-dependent partonic quenching,
accompanied by a greater relative contribution
from surviving jet-driven nonflow correlations.
Inclusive neutral-meson measurements at
$\sqrt{s_{NN}}=8.16$~TeV show $R_{\rm pPb}$
compatible with unity above approximately
$10$~GeV/$c$~\cite{ALICE:2021est}, providing
complementary context, although these measurements
do not isolate the ultra-central event class.
Together, these observations constrain the
collision-environment dependence of the relative
charm response and effective transition momentum.

Taken together, the Pb+Pb, p+Pb, and Au+Au results
establish a stringent benchmark for microscopic
heavy-quark transport descriptions. 
A successful description must reproduce the approximate
$D^0$ $v_2$--$v_3$ closure in Pb+Pb collisions at
$\sqrt{s_{NN}}=5.02$~TeV, the systematic charm--light
response difference revealed by the null test, and its
evolution across centrality, collision system, and
beam energy. It must also account for the comparable charm 
and light-meson transition scales in Pb+Pb, the consistency
of the p+Pb $D^0$ measurements with the substantially
lower light-meson transition scale, and the near-unity
charm attenuation rescaling observed in Au+Au at
$\sqrt{s_{NN}}=200$~GeV.
These simultaneous constraints help distinguish the
common medium response from the additional charm
response, reducing ambiguities in the interpretation
of heavy-quark transport.

%
In summary, the independently constrained light-flavor
response provides a direct null test of $D^0$ anisotropy
scaling. The transformed $D^0$ $v_2$ and $v_3$ exhibit
approximate inter-harmonic closure but remain systematically
displaced from the kaon-anchored scaling function,
establishing a distinct charm scaling violation.
The common response is recovered primarily through
the charm-specific attenuation rescaling $k_c$,
together with a small low-$p_T$ contribution
through the established hadronic-rescattering
response channel, $\zeta_{\rm hs}\simeq0.06$.
The coefficient $k_c$ decreases from approximately
$2.0$ in central and mid-central Pb+Pb collisions
to $1.27$ in the 30--50\% interval, while the
$D^0$ $v_2$ measurements in p+Pb and Au+Au give
$k_c\simeq1.5$ and $k_c\simeq1$, respectively.
 The light-flavor attenuation coefficient
also differs between RHIC and the LHC, distinguishing
the common medium response from the additional charm
response. Despite these differences, the charm and light-meson
responses are consistent with a common transition scale
of $4.5$--$5.0$~GeV/$c$ in Pb+Pb, while the p+Pb
$D^0$ measurements remain consistent with the
substantially lower transition scale independently
established for light mesons. The combined harmonic, momentum,
centrality, collision-system, and beam-energy
constraints establish a differential experimental
benchmark for heavy-quark transport and medium evolution.
%
\section*{Acknowledgement}
This work is supported in part by the U.S. Department of Energy, Office of Science, 
Office of Nuclear Physics, Grant Number DE-SC0024602.
\bibliography{pid-refs-D0}
%
\end{document}